\documentstyle[12pt,proc,psfig]{article}
\begin{document}
\centerline{\tenbf SUPERSYMMETRY FOR FLAVORS} 
\vspace{0.8cm}
\centerline{\tenrm Chun Liu}
\baselineskip=13pt
\centerline{\tenit Institute of Theoretical Physics,
Chinese Academy of Sciences} 
\baselineskip=12pt
\centerline{\tenit Beijing 100080, China}
\vspace{0.9cm}
\abstracts{An understanding of the lepton masses within the framework of
SUSY is presented. A family symmetry is introduced. Sneutrino VEV breaks
this symmetry. The tau mass is due to the Higgs VEV, and muon mass purely
from the sneutrino VEV. A viable model is constructed, which predicts
($1-10$) MeV $\nu_{\tau}$.}
\vfil
\twelverm
\baselineskip=14pt

\section{Introduction}

The naturalness of the standard model (SM) \cite{1} may require the
existence of low energy supersymmetry (SUSY) \cite{2}. The flavor puzzle,
namely the fermion masses, mixings and CP violation, of the SM needs new
physics to be understood. It would be nice if SUSY also provides (partial)
understanding to the flavor problem \cite{3}.

Let us look at the fermion mass pattern. The fact is that the third
generation is much heavier than the second generation which is also much
heavier than the first. Does this imply a family symmetry? We assume that
the answer is yes. Let us consider the charged leptons. By assuming a $Z_3$
cyclic symmetry among the SU(2) doublets $L_i$ ($i=1,2,3$) of the three
generations \cite{3}, the Yukawa interactions result in a democratic mass
matrix. This mass matrix is of rank $1$. Therefore only the tau lepton gets
mass, the muon and electron are still massless.

The essential point is how the family symmetry breaks. Naively the symmetry
breaking can be achieved by introducing more Higgs fields. We consider this
problem within SUSY. We have observed that SUSY naturally provides Higgs
like fields, which are the scalar neutrinos. Furthermore, if the vacuum
expectation values (VEVs) of the sneutrinos are non-vanishing, $v_i\neq 0$,
the R-parity violating interactions $L_iL_jE^c_k$, with $E^c_k$ denoting the
anti-particle superfields of the SU(2) singlet leptons, contribute to the
fermion masses, in addition to the Yukawa interactions. This made us to
propose that the family symmetry is broken by the sneutrino VEVs \cite{3}.

A remark should be made here. The sneutrino VEV is not enough to the
electroweak symmetry breaking (EWSB), the Higgs fields are still necessary.
As will be seen, $v_i$ is typically ($5-10$) GeV. Such a VEV however, is too
large to accommodate the neutrino oscillation data. The point we made is
that such a sneutrino VEV breaks a family symmetry which can be useful for
the understanding of the charged lepton masses.

Let us focus on the lepton sector. The $Z_3$ family symmetry results in that
Yukawa interactions only give $\tau$ lepton mass, $m_{\tau}\simeq yv_d$,
where $y$ is the coupling constant, and the Higgs VEV $v_d\sim v_u\sim 100$
GeV. Taking $y\sim 10^{-2}$ gives realistic $m_{\tau}$. The muon mass is due
to the family symmetry breaking $v_i\neq 0$. To make the model
phenomenologically acceptable, we found $v_3\sim 10$ GeV, and the trilinear
R-parity violating couplings $\lambda\sim 10^{-2}$. The muon mass is then
$m_{\mu}\simeq\lambda v_3\simeq 100$ MeV.

\section{Challenging Problems}

Immediately the following questions are raised to challenge the above
described scenario:

1. Is $10$ GeV $v_3$ safe? Namely do we have unacceptable Majoron?

2. Is $m_{\nu_{\tau}}$ too large? Roughly it is expected to be
$g_2^2v_3^2/M_Z\sim 100$ MeV $-1$ GeV, with $g_2$ being the coupling of the
SU(2) interaction.

3. Is $\lambda\sim 10^{-2}$ contradict with the experimental data on the
rare decays $\mu\to 3e$ and $\mu\to e\gamma$?

To get rid of the above difficulties, the following things are assumed
respectively. First, a large (weak scale) mixing mass term of the Higgs
scalar and slepton $\tilde{L}_3$, $B\mu_3\tilde{H}_u\tilde{L}_3$ is
introduced. This term breaks SUSY softly. Because it breaks the lepton
number explicitly, no massless Majoron which would be the Goldstone particle
corresponding to the spontaneous lepton number violation, appears. Note that
the trilinear R-parity violating interactions themselves are not enough to
keep the would be Majoron from being light. Second, in the superpotential,
the bilinear mixing mass terms, like $\mu_3H_uL_3$ and $\mu H_uH_d$ should
be small. We take $\mu_3=0$ and $\mu=0$. In this way, $m_{\nu_{\tau}}=0$ at
tree level. This may avoid the difficulty of Question 2. Note that the EWSB
is achieved by an alternative superpotential. Third, the family symmetry
$Z_3$ in the ordinary Yukawa and trilinear R-parity violating interactions is
adopted. Because of this, $\mu\to 3e$ and $\mu\to e\gamma$ do not occur.
This symmetry can be regarded as accidental at low energy.

The relevant superpotential is 
\begin{equation}
\begin{array}{lll}
{\cal W} & = & \displaystyle y_j(\sum_{i}^{3}L_i)H_dE^c_j
+\lambda_j(L_1L_2+L_2L_3+L_3L_1)E^c_j \\[3mm] 
&  & \displaystyle +\lambda^{\prime}X(H_uH_d-\mu^2)~, \label{1}
\end{array}
\end{equation}
where the last term is for the EWSB with $\lambda^{\prime}$ being the
coupling constant, $X$ a singlet superfield and $\mu$ the weak scale. The
baryon number conservation is assumed \cite{4}. Eq. (\ref{1}) reproduces the
lepton masses as we have planned. It would be easier to work in the mass
eigenstates of Yukawa interactions, 
\begin{equation}
{\cal W}=\displaystyle y^{\tau}L_{\tau}H_dE^c_{\tau}
+L_eL_{\mu}(\lambda_{\mu}E^c_{\mu}+\lambda_{\tau}E^c_{\tau})
+\lambda^{\prime}X(H_uH_d-\mu^2)~,  \label{2}
\end{equation}
where the left-handed leptons are 
\begin{equation}
\begin{array}{lll}
\tau_L & = & \displaystyle\frac{1}{\sqrt{3}}(e_1+e_2+ e^{\prime}_3)~,
\\[3mm]
\mu_L & = & \displaystyle\frac{1}{\sqrt{2}}(e_1-e_2)~, \\[3mm] 
e_L & = & \displaystyle\frac{1}{\sqrt{6}}(e_1+e_2-2e^{\prime}_3)~,
\end{array}
\end{equation}
with ($\nu_i$, $e_i$) being the fermionic component of $L_i$.
($\nu_i^{\prime}$, $e_i^{\prime}$) means the physical leptons after
considering the mixing with the neutralinos and charginos. From Eq.
(\ref{2}), we see that $\mu\to 3e$ does not occur. Instead, the rare
decays $\tau\to 2e\mu$ and $3\mu$ have branching ratios $10^{-7}$ if
$m_{\tilde{\nu}_i}\simeq 100$ GeV.

At the quantum level, a comparatively large neutrino mass is inevitably
induced due to the large lepton number violating effect in $B\mu_3$ mass. It
occurs at the one-loop level with a Zino exchange. Therefore
$m_{\nu_{\tau}}\neq 0$ at one loop which will be studied further. Is it
natural in a theory in which $B\mu_3$ is large, $\mu_3$ is vanishingly small
and $m_{\nu_{\tau}}$ is consistent with experiment?

\section{A Model of GMSB}

With the framework of gauge mediated SUSY breaking (GMSB) \cite{5}, the
scenario asked in the last question can be realized naturally \cite{6}.
Lepton number violation is introduced originally in the messenger sector. It
is then communicated to the SM sector including the related soft SUSY
breaking terms. We will make use of the observation of the $\mu$-problem in
GMSB \cite{7}. It was noted that both $\mu$ term and its corresponding soft
breaking $B\mu$ term can be generated at one loop. Either $\mu$ is at the
weak scale and $B\mu$ is unnaturally large, or $B\mu$ is at the weak scale
and $\mu$ is very small. This is not a problem in our model, because the
EWSB given in Eq. (\ref{1}) does not need the $\mu$ term. However, we apply
similar observation to the discussion of the mixing of $H_u$ and $L_3$.

The messengers are introduced as follows with the
SU(3)$\times$SU(2)$\times$U(1) quantum numbers,
\begin{equation}
S,S^{\prime }=(1,2,-1)~,~~~~~\bar{S},\bar{S^{\prime }}=(1,2,1)~,
\end{equation}
and 
\begin{equation}
T,T^{\prime }=(3,1,-2/3)~,~~~~~\bar{T},\bar{T^{\prime }}=(\bar{3},1,2/3)~.
\end{equation}
Two more gauge singlets are introduced $Y$ for the SUSY breaking and $V$ for
the lepton number violation. The superpotential is then
\begin{equation}
{\cal W}_{\rm total}={\cal W}+{\cal W}_1+{\cal W}_2~,
\end{equation}
where  
\begin{equation}
\begin{array}{lll}
{\cal W}_1 & = & m_1(\bar{S^{\prime }}S+S^{\prime }\bar{S})+m_2(
\bar{T^{\prime }}T+T^{\prime }\bar{T})+m_3S\bar{S}+m_4T\bar{T}+m_5V^2 \\[3mm]
&  & +Y(\lambda _1S\bar{S}+\lambda _2T\bar{T}+\lambda _3V^2-
\mu_1^2)~,\nonumber
\end{array}
\end{equation}
\begin{equation}
{\cal W}_2=V(\lambda _5H_uS+\lambda _6L_3\bar{S})~,
\end{equation}
with $\mu_1$ being the SUSY breaking scale.  The lepton number violation lies
in ${\cal W}_2$.  By integrating out the heavy messengers, the effective
Lagrangian related to lepton number violation is
\begin{equation}
{\cal L}_{{\rm eff}}^{\not{L}}=\mu _3 L_{3}H_u|_{\theta \theta }+
B\mu _3\tilde{L_3}\tilde{H_u}+{\rm h.c.}~,
\end{equation}
where
\begin{equation}
\mu _3\simeq\displaystyle \frac{\lambda _5\lambda _6}{16\pi ^2}
\frac{\mu _1^2}{m_3}~,~~
B\mu _3\simeq\displaystyle\frac{\lambda _5\lambda _6}{16\pi ^2}
\left( \frac{\mu _1^2}{m_3}\right) ^2~.
\end{equation}
Or
\begin{equation}
B\mu _3=\mu _3 \frac{\mu _1^2}{m_3}~.
\end{equation}
This is what we needed.  $\mu_1^2/m_3$ is usually around $100$ TeV.  $\mu_3$
is very small if $B\mu_3$ is taken to be the weak scale.

From the scalar potential, the sneutrino VEV is obtained as 
\begin{equation}
v_1=0~,~~ v_2=0~,~~ v_3=-\frac{B\mu_3v_u}{M_A^2+\frac{1}{2}M_Z^2\cos2\beta}~.
\end{equation}
Numerically, $v_3\sim 10$ GeV if $M_A\sim 300$ GeV and
$B\mu_3\sim (60~{\rm GeV})^2$.  Nonzero $v_3$ implies the mixing between
neutrino $\nu_3$ and the neutralinos.  In addition, $B\mu_3$ causes
comparatively large $\nu_3$-Higgsino mixing at one loop,
\begin{equation}
m_{3H}\simeq \mu_3 +\frac{g_2^2}{16\pi^2}\frac{B\mu_3}{M_{\tilde{Z}}}
\sim 0.04-0.1~ {\rm GeV}~.
\end{equation}

The $\tau$ neutrino mass should be obtained from the full neutralino mass
matrix,
\begin{equation}
-i(\nu _3~~H_d^0~~H_u^0~~\tilde{Z}~~X)\left( 
\begin{array}{ccccc}
0 & 0 & m_{3H} & av_3 & 0 \\ 
0 & 0 & 0 & av_d & \lambda^{\prime }v_u \\ 
m_{3H} & 0 & 0 & -av_u & \lambda ^{\prime }v_d \\ 
av_3 & av_d & -av_u & M_{\tilde{Z}} & 0 \\ 
0 & \lambda ^{\prime }v_u & \lambda ^{\prime }v_d & 0 & 0
\end{array}
\right) \left( 
\begin{array}{c}
\nu _3 \\ H_d^0 \\ H_u^0 \\ \tilde{Z} \\ X
\end{array}
\right) +{\rm h.c.},
\end{equation}
with $a=\displaystyle(\frac{g_1^2+g_2^2}2)^{1/2}$. It gives
\begin{equation}
m_{\nu _\tau }\simeq \frac{m_{3H}v_3}{M_Z}\sim (1-10)~{\rm MeV}~.
\end{equation}

This heavy $\nu_\tau$ can decay to $e^+e^-\nu_e$.  We later noted that
$\nu_\tau$ can decay to gravitino $+$ photon with a longer lifetime.  In
writing down the expression of the physical $\nu_\tau$ and $\tau$ states
in Refs. \cite{3} and \cite{6}, the gaugino masses were neglected.  In
fact, $\nu_3$ mixes with photino.

\section{Discussion}

The idea about fermion masses presented in this model essentially depends
on SUSY.  A model of large sneutrino VEV exists.  ($1-10$) MeV $\nu_\tau$
is a consequence of this VEV.  It should be noted that $L_3$ and $H_d$
appear in the superpotential in different ways, so that the VEV cannot be
rotated away through redefining Higgs superfield.

The atmospheric neutrino anomaly must be explained by introducing a
sterile neutrino which is also necessary from the constraint of the
Big-Bang Neucleosynthesis.  The extension of this idea to the quark
sector can be found in Ref. \cite{8}.

The $B\mu_3$ term breaks the family symmetry explicitly.  It would be
more appealing if the family symmetry breaking is spontaneous in some
clever model.

\section{Acknowledgments}

I would like to thank Profs. Dongsheng Du and H.S. Song for collaborations.

\end{document}